\documentclass{jetpl}

\usepackage{ucs} 
\usepackage[koi8-r]{inputenc}
\usepackage[english,russian]{babel}

\twocolumn

\rus

\title{ Torsion-rotational transitions in methanol as a probe of fundamental physical constants -- electron and proton masses 
}

\rtitle{ Torsion-rotational transitions in methanol\ldots}

\sodtitle{Torsion-rotational transitions in methanol as a probe of  fundamental physical constants -- electron and proton masses}

\author{J.\,S.\,Vorotyntseva\/\thanks{yuvorotynceva@yandex.ru},
        S.\,A.\,Levshakov}

\rauthor{J.\,S.\,Vorotyntseva, S.\,A.\,Levshakov}

\sodauthor{Vorotyntseva, Levshakov}

\address{Ioffe Institute, Politechnicheskaya 26,
St. Petersburg 194021, Russia}

\dates{}{*}

\abstract{We report on the using of torsion-rotational transitions in the CH$_3$OH and $^{13}$CH$_3$OH molecules to evaluate possible variations of the physical constant $\mu=m_e/m_p$ -- the electron-to-proton mass ratio -- from spectral observations of emission lines detected in the microwave range towards the dense molecular cloud Orion-KL. An estimate of the upper limit on the relative changes in $\mu$ is obtained by two independent ways -- with $^{13}$CH$_3$OH lines and with the combination of $^{13}$CH$_3$OH and CH$_3$OH lines. The calculated upper limit $\Delta\mu/\mu<1.1\times10^{-8}$ (1$\sigma$) is in line with the most stringent constraints on the variability of fundamental physical constants established by other astrophysical methods.}

\PACS{???}

\begin{document}

\maketitle

The existence of Dark Matter in the Universe follows from a number of observational facts, including flat rotation curves of galaxies at large galactocentric distances, gravitational lensing of cosmologically distant objects and the large-scale structure of the spatial distribution of galaxies ~\cite{bib:Ber}. To explain the nature of Dark Matter, various models have been considered, including phenomena extending the Standard Model of particle physics (SM). Several of these models assume the existence of hypothetical scalar fields interacting with the baryonic component of an ordinary substance ~\cite{bib:OR2010, bib:FDA, bib:Alex}. The result of such interactions can lead to the space-time variations of dimensionless physical constants, such as the fine structure constant $\alpha$ and the electron-to-proton mass ratio $\mu$ ~\~\cite{bib:Uz}-\cite{bib:Oli}.
Therefore, the new theories can be tested experimentally using the relative measurements of $\mu=m_e/m_p$. It should be noted that the mass of the electron $m_e$ is directly related to Higgs-like scalar fields, while the main contribution to the mass of the proton $m_p$ comes from the binding energy of quarks.
 
Electron-vibro-rotational transitions in molecular spectra have a specific dependence on $\mu$, individual for each transition ~\cite{bib:VL,bib:KL}. The reaction of the transition to a change in $\mu$ is characterized by a dimensionless sensitivity coefficient $Q_\mu$, which is defined as
\begin{equation}
Q_\mu = \frac{df/f}{d\mu/\mu},
\label{eq:1}
\end{equation}
here $df/f$ is the relative frequency offset, and $d\mu/\mu$ is given by
\begin{equation}
\frac{\Delta\mu}{\mu}= \frac{\mu_{obs}-\mu_{lab}}{\mu_{lab}},
\label{eq:2}
\end{equation}
where $\mu_{obs}$, $\mu_{lab}$ are astronomical and laboratory values of
$\mu$, correspondingly. At the same time, the sensitivity coefficients can take different signs, which leads to an increase or decrease in the observed frequency compared to its laboratory value.

The most stringent limits on $\mu$-variations at large redshifts $z$ were obtained from extragalactic observations towards the quasar J1443+2724 ($z$ = 4.22). From the analysis of the Lyman and Werner absorption lines of molecular hydrogen 
H$_2$ the upper limit on $\Delta\mu/\mu<8\times10^{-6}$ was obtained
~\cite{bib:UB}. In the Milky Way, the most stringent upper limits were established from observations of the 23 GHz inversion transition in ammonia NH$_3$, which has a 
sensitivity coefficient $Q_\mu = 4.46$ ~\cite{bib:FK}, compared to the pure
rotational transitions in HC$_3$N, HC$_5$N, and HC$_7$N, which have $Q_\mu = 1$:
$\Delta\mu/\mu<7\times10^{-9}$ ~\cite{bib:L13}.
Independent estimates from observations of the thermal emission lines of methanol CH$_3$OH
in the core of the molecular cloud L1498 lead to a value of 
$\Delta\mu/\mu<2\times10^{-8}$ ~\cite{bib:D}.
Similar constraints follow from the measurements of the radial velocities 
of methanol maser lines: $\Delta\mu/\mu<2\times10^{-8}$ ~\cite{bib:L22} and  
$\Delta\mu/\mu<2.7\times10^{-8}$ ~\cite{bib:E11}.
All estimates of $\Delta\mu/\mu$ are given at a 1$\sigma$ confidence level.

It should be noted that in previous works
methanol isotopologues
have not been used extensively. The first estimates of the upper limit on $\mu$-variations
were obtained from the TMRT 65-m telescope ~\cite{bib:Wu} observations of the
thermal emission lines of $^{13}$CH$_3$OH
in the star-forming region NGC 6334I: $\Delta\mu/\mu<3\times10^{-8}$.
~\cite{bib:VKL}.

In the molecules with hindered internal motion, enhanced sensitivity coefficients
$Q_\mu$ are inherent for tunneling transitions, since the tunneling probability depends
exponentially on the masses of tunneling particles ~\cite{bib:FK,bib:LKR, bib:J11}.
The most promising molecule for these studies is methanol (CH$_3$OH),
where the methyl group CH$_3$ can make torsional vibrations relative to the hydroxyl group OH.
In this case, the hydrogen atom of the hydroxyl group can be located in three
possible positions with equal energies, and to move from one configuration to another,
it must pass through a potential barrier caused by the three hydrogen atoms 
of the methyl group. So, there is an internal hindered motion 
of the hydrogen atom relative to the methyl group.

The sensitivity coefficients $Q_\mu$ for methanol were calculated by two 
independent methods in 2011 ~\cite{bib:LKR, bib:J11}. The results 
show that low-frequency (in the range of 1--50 GHz) transitions have 
high values of $Q_\mu$ with different signs: $-17 \leq Q_\mu \leq +43$, which,  compared to  the sensitivity coefficients of the molecular hydrogen H$_2$ lines ($|Q_\mu|\sim 10^{-2}$) give a gain in the limiting estimates of $\Delta\mu/\mu$ by a factor of more than 1000.

In our previous work ~\cite{bib:VKL}, a list of molecules with the enhanced
sensitivity coefficients was extended due to the methanol isotopologues -- $^{13}$CH$_3$OH with $-32 \leq Q_\mu \leq +78$, and 
CH$_3$$^{18}$OH with $-109 \leq Q_\mu \leq +33$.

Turning to the practical measurements of $\Delta\mu/\mu$, we note that to estimate this value,
pairs of molecular lines with different coefficients  
$Q_{\mu,1}$ and $Q_{\mu,2}$ are used ~\cite{bib:LKR}:
 \begin{equation}
\frac{\Delta\mu}{\mu}= \frac{V_1-V_2}{c(Q_{\mu,2}-Q_{\mu,1})},
\label{eq:3}
\end{equation}
where $V_1$ and $V_2$ are the measured radial velocities
of these lines, and $c$ is the speed of light. The conversion from the frequency scale $f$
to the velocite scale $V$
is made by the radio astronomical definition
$V/c = (f_{lab}-f_{obs})/f_{lab}$.

The accuracy of the $\Delta\mu/\mu$ measurements is due to the influence of various factors.
Uncertainties in laboratory frequencies and line centers in astronomical spectra 
are the main sources of errors. In addition, there are systematic errors which can be estimated from observations of different objects in different molecular transitions.
Transitions in methanol isotopologues have high sensitivity coefficients of both signs, i.e., they are the most suitable candidates for such studies.

High precision measurements can be carried out from high-resolution spectra, which were recently obtained
for the Orion-KL molecular cloud ~\cite{bib:Liu}. The published 
spectra contain lines of methanol CH$_3$OH and its two isotopologues --
$^{13}$CH$_3$OH and CH$_3$$^{18}$OH. The CH$_3$$^{18}$OH lines are found to be quite
weak, and have large errors in radial velocities. However, the $^{13}$CH$_3$OH emission spectra
display lines with higher intensities, and their positions are determined
quite accurately (with errors of 100 m~s$^{-1}$, which are acceptable
for our purposes). Thus, it becomes possible to estimate $\Delta\mu/\mu$
independently~--
with only the lines of $^{13}$CH$_3$OH and in combination with the transition 
$J_{K_u}\to J_{K_l} = 15_2-15_1 E$ in CH$_3$OH (see Table 1 below).

%-------------------------Table 1
\begin{table*}
\centering
\caption\centering{\bf{Table 1. Selected $^{13}$CH$_3$OH and CH$_3$OH transitions
in Orion-KL~\cite{bib:Liu}. \\
Given in parentheses are errors in the last digits.}}
\label{T}
\begin{tabular}{r c c c c c }
\hline
\multicolumn{1}{c}{Molecule}&Transition &$f_{lab}$& $\Delta v_{\rm D}$& $V_{\rm LSR}$& ${Q_\mu}$\\
\multicolumn{1}{c}{}&$J_{K_u}-J_{K_l}$&[MHz]& [km s$^{-1}$]& [km s$^{-1}$]&\\
\hline
{$^{13}$CH$_3$OH}&$6_{2}-5_{3}$A$^{-}$& 27992.990 & 2.3(3)& 6.9(1)& 16.3\\ 
{$^{13}$CH$_3$OH}&$9_{2}-9_{1}$E& 27581.630 & 3(1)& 6.8(1)& $-$14.7\\ 
{CH$_3$OH}&$15_{2}-15_{1}$E& 28905.812 & 2.9(1)& 6.6(1)& $-13.3^1$\\ 
\hline

\multicolumn{6}{l}{\footnotesize $^1$ -- The sensitivity coefficient was calculated in the present work.}\\
\end{tabular}
\end{table*}

We selected from the published spectra the pairs of molecular  transitions with approximately equal values of the Doppler
widths $\Delta v_D$, such that the difference in the sensitivity coefficients $\Delta Q_\mu$ is maximized. The selected lines and their parameters are given
in Table 1. The first column lists the transition described by two quantum numbers
-- the total angular momentum $J$ and its projection $K$ on the principal axis of the molecule -- for the upper
($u$) and lower ($l$) levels, the second column displays the transition frequency, the third and fourth columns show the Doppler width and the Local Standard of Rest radial velocity, $V_{LSR}$, respectively. 
The fifth
column contains the sensitivity coefficient $Q_\mu$, taken from ~\cite{bib:VKL}.
The calculation of $Q_\mu$ for methanol CH$_3$OH was performed in this work using our previously
developed method ~\cite{bib:LKR}.
As can be seen from Table 1, the difference in the sensitivity coefficients is $\Delta Q_\mu \approx 30$,
which provides a confident estimate on the $\mu$-variance.

 Using Eq. (\ref{eq:3}) for the transitions in $^{13}$CH$_3$OH, we obtain $\Delta\mu/\mu=(-1.1\pm1.5)\times10^{-8}$, which corresponds to the 
upper limit on $\Delta\mu/\mu<1.5\times10^{-8}$.
A similar calculation for the combination of the $6_2-5_3 A^-$ line of $^{13}$CH$_3$OH
and the $15_2-15_1 E$ line of CH$_3$OH gives $\Delta\mu/\mu=(-3.4\pm1.6)\times10^{-8}$,
and the upper limit on $\Delta\mu/\mu<1.6\times10^{-8}$.
This yields the average value of $\langle \Delta\mu/\mu \rangle = (-2.3\pm1.1)\times10^{-8}$,
and the corresponding upper limit on the changes in $\mu$ -- $\Delta\mu/\mu<1.1\times10^{-8}$.
This upper limit agrees well with previously obtained constraints from Galactic
CH$_3$OH ~\cite{bib:D, bib:L22} and $^{13}$CH$_3$OH ~\cite{bib:VKL} observations.

The results of these investigations do not indicate any significant
systematic errors in the estimates of $\Delta\mu/\mu$.
It follows that the assumed effects of Higgs-like 
scalar fields on the masses of elementary particles do not exceed 
the 10$^{-8}$ level in the disk of the Galaxy. 
This upper limit, 10$^{-8}$, also coincides with a limit on the influence
of the hypothetical fifth force on hadronic interactions ~\cite{bib:SUK},
so it can be considered as the most robust 
at the current time.

\textbf{FUNDING}\\
This work was supported by ongoing institutional funding and was carried out within the framework of the topic of the State Assignment
of the Ioffe Institute number FFUG-2024-0002. No additional grants to carry out or direct this particular research were obtained.

\textbf{CONFLICT OF INTEREST}\\
The authors declare no conflict of interest.

\vfill\eject


\begin{thebibliography}{16}


\bibitem{bib:Ber}
G. Bertone, D. Hooper, Rev. Mod. Phys. {\bf 90}, 045002 (2018).


\bibitem{bib:OR2010}
R.\, Onofrio, Phys. Rev. D {\bf 82}, 065008 (2010).


\bibitem{bib:FDA}
F.\, D. Albareti et al., Phys. Rev. D {\bf 95}, 044030 (2017).


\bibitem{bib:Alex}
S.\, Alexander et al., CQD {\bf 33}, 14LT01 (2016).

\bibitem{bib:Uz}
J.-P. Uzan, Living Rev. Phys. {\bf 14}, 2 (2011).

\bibitem{bib:Bek}
J. D.  Bekenstein, Phys. Rev. D {\bf 25} 1527 (1982).

\bibitem{bib:Bra}
P. Brax, Phys. Rev. D {\bf 90} 023505 (2014).

\bibitem{bib:Oli}
K. A. Olive, M. Pospelov, Phys. Rev. D {\bf 77}, 043524 (2008).


\bibitem{bib:Th}
R.\, I. Thompson, Astrophys. Lett. {\bf 16}, 3 (1975).


\bibitem{bib:VL}
D.\, A. Varshalovich, S.\, A. Levshakov, J. Exp. Theor. Phys. Lett. {\bf 58}, 237 (1993).


\bibitem{bib:KL}
M.\, G. Kozlov, S.\, A. Levshakov, Ann. Phys. {\bf 525}, 452 (2013).


\bibitem{bib:UB}
J.\, Bagdonaite et al., Phys. Rev. Lett. {\bf 114}, 071301 (2015).


\bibitem{bib:FK}
V.\, V. Flambaum, M.\, G. Kozlov, Phys. Rev. Lett. {\bf 98}, 240801 (2007).


\bibitem{bib:L13}
S.\,A. Levshakov et al., Mem. S. A. It., {\bf 85}, 90 (2014).


\bibitem{bib:D}
M.\, Dapr\`a et al., MNRAS {\bf 472}, 4434 (2017).

\bibitem{bib:L22}
S.\, A. Levshakov et al., MNRAS {\bf 511}, 413 (2022).

\bibitem{bib:E11}
S. Ellingsen, M. Voronkov, S. Breen, Phys. Rev. Lett {\bf 107}, 270801 (2011).

\bibitem{bib:Wu}
J.\, H. Wu et al., ApJS {\bf 265}, 49 (2023).


\bibitem{bib:VKL}
J.\, S. Vorotyntseva, M.\, G. Kozlov, S.\, A. Levshakov, MNRAS {\bf 527}, 2750 (2024).


\bibitem{bib:LKR}
S.\,A. Levshakov, M.\,G. Kozlov, D. Reimers, ApJ {\bf 738}, 26 (2011).


\bibitem{bib:J11}
P.\, Jansen et al., Phys. Rev. Lett. {\bf 106}, 100801 (2011).


\bibitem{bib:Liu}
X.\, Liu et al., ApJS, {\bf 106}, 19 (2024).

\bibitem{bib:SUK}
E. J. Salumbides, W. Ubachs, V. I. Korobov, J. Mol. Spec. {300}, 65 (2014).

\end{thebibliography}
\end{document}